\documentclass[twocolumn,aps,prb,showpacs]{revtex4} 

\usepackage{amsmath}
\usepackage{epsfig}

\newcommand{\ave}[1]{\ensuremath{\langle #1 \rangle}}
\newcommand{\D}{\Delta}
\newcommand{\de}{\delta}
\newcommand{\e}{\mathrm{e}}
\newcommand{\A}{\alpha}
\newcommand{\be}{\begin{equation}}
\newcommand{\ee}{\end{equation}}
\newcommand{\ba}{\begin{eqnarray}}
\newcommand{\ea}{\end{eqnarray}}

\begin{document}

\title{The conversion of phase to amplitude fluctuations of a light beam by an optical cavity}

\author{Alessandro S. Villar}
\email{alessandro.villar@uibk.ac.at}
\altaffiliation[Former address: ]{Instituto de F\'\i sica, Universidade de S\~ao Paulo,
Caixa Postal 66318, 05315-970 S\~ao Paulo, SP, Brazil.}

\affiliation{Institut f\"ur Experimentalphysik, Universit\"at Innsbruck, Technikerstrasse 25/4, 6020 Innsbruck, Austria,
Institute for Quantum Optics and Quantum Information, Austrian Academy of Sciences, 6020 Innsbruck, Austria}

\begin{abstract}
Very low intensity and phase fluctuations are present in a bright light field such as a laser beam. These subtle quantum fluctuations may be used to encode quantum information. Although intensity is easily measured with common photodetectors, accessing the phase information requires interference experiments. We introduce one such technique, the rotation of the noise ellipse of light, which employs an optical cavity to achieve the conversion of phase to intensity fluctuations. We describe the quantum noise of light and how it can be manipulated by employing an optical resonance technique and compare it to similar techniques, such as Pound-Drever-Hall laser stabilization and homodyne detection.
\end{abstract}


\maketitle

\section{INTRODUCTION}

Light fields are an important tool in the field of quantum information.\cite{nielsenchuangQCQI,reviewbraunsteinvanloock_rmp05} The intensity and the phase of bright light beams possess quantum properties similar to the position and momentum of a quantum harmonic oscillator:\cite{fabreILOQ} it is impossible to know both at the same time with arbitrary precision because they must satisfy the uncertainty principle. The Einstein-Podolsky-Rosen\cite{epr_physrev35} paradox can be realized with these observables, giving rise to entangled beams. Entanglement is important because it is the primary resource in quantum computation and information.\cite{jozsa_procrsoca03,forcer_quantuminformcompu03} That is, quantum information can be encoded in the intensity and phase of light.

\begin{figure}[ht]
\centering
\epsfig{file=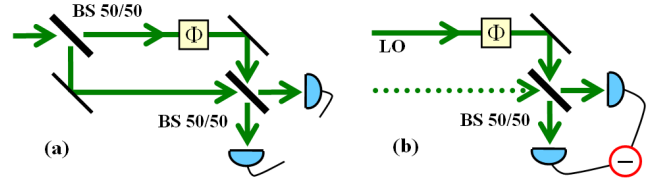,scale=0.35}
\caption{(a) In the Mach-Zehnder interferometer the relative phase $\Phi$ between the two possible paths determines the intensity difference between the two output ports. (b) Homodyne detection is performed by interfering in a balanced beam splitter (BS 50/50) the field to be measured (dotted arrow) and a strong local oscillator field (LO, solid arrow). The subtraction of the photocurrents gives information about the phase fluctuations of the target beam when the mean relative phase is $\Phi=\pi/2$.}
\label{homodyne}
\end{figure}

To employ this entanglement in practical implementations, the two observables must be accessible to measurement. Light intensity is easily measured by photodetectors. In contrast, phase information requires an interference experiment and must be converted into amplitude information as can be understood by considering the Mach-Zehnder interferometer\cite{machzehnder_ajp95} for example (see Fig.~\ref{homodyne}(a)). If the phase difference $\Phi$ between the two optical paths is zero (or $\pi$), light exits the interferometer through one (or another) of two output ports; if $\Phi=\pi/2$, then the beam is equally divided between them. Thus, by measuring the intensity difference between the output ports, it is possible to infer the relative phase between the two paths. 

In this example, only the mean relative phase was considered, but a similar situation holds for fast phase fluctuations as well. If another intense light beam is used with a stable phase relative to the beam to be measured, it is possible to acquire phase information about the target beam by their interference (Fig.~\ref{homodyne}(b)). It suffices to choose their relative mean phase as $\pi/2$, as in the Mach-Zehnder interferometer example, such that the two outputs are balanced. Then a small relative phase fluctuation between the two fields is optimally converted to a fluctuation in the difference of intensities. This technique is called homodyne detection, the most commonly employed method of measuring the quantum phase fluctuations of light.\cite{bachorGEQO} The practical implementation of a homodyne technique might be difficult in some situations, including the establishment of an auxiliary beam that matches the field to be measured (phase reference, frequency, and spectral profile) and the requirement of a much higher intensity which might saturate the photodetectors.\cite{prlentangtwinopo}

One way to circumvent these difficulties lies in a powerful method used for laser frequency stabilization, the Pound-Drever-Hall technique.\cite{pdh_ajp01,pdhoriginal_applphysb83} The original idea behind it is to modulate the beam phase, which may be interpreted as the creation of ``sidebands'' in its spectrum, and compare the light beam frequency to an optical cavity resonance frequency. The cavity, as for every resonance, has a dispersive character, dephasing each frequency component by different amounts. By analyzing the field reflected by the cavity, it is possible to obtain with great precision the detuning (frequency difference) between the laser and cavity based on the dephasing between carrier and sidebands. This information is then fed back to the laser or the cavity in order to lock their frequencies to each other. The action of the cavity can be interpreted as converting a small amount of the incident beam phase modulation into amplitude modulation of the reflected beam, which can then be recorded by a photodetector. At perfect resonance, no conversion occurs; on different sides of it, the conversions have opposite signs; it is thus possible to know if the laser frequency is equal to, higher, or lower than the cavity resonance frequency.

This technique can be adapted to measure the quantum sidebands of light beams.\cite{levenson_pra85,galatola_optcomm91} The requirement is that the conversion be efficient; that is, all the original quantum phase fluctuations of the incident beam be converted into amplitude fluctuations in the reflected one. This conversion is possible if certain conditions are met. It is necessary that the cavity resonance be sufficiently narrow to clearly distinguish the carrier from the sidebands. Then the carrier acts as a local oscillator and the sidebands as a faint light field. The cavity introduces a relative phase shift between them, accomplishing, as in homodyne detection, the full conversion of phase to amplitude noise. The frequency, spectral profile, phase reference, and spatial mode are automatically matched between the two fields, because they are different spectral components of a single beam. No extra noise is added by the local oscillator. Because of the picture of quantum noise as a noise ellipse,\cite{wallsmilburnQO} we call this technique the ``rotation of the noise ellipse of light.'' 

In Sec.~\ref{seclightnoise} the basic concepts of the quantum properties of laser-like beams are presented. Section~\ref{secoticalcavity} introduces the optical cavity, and its effect on the light field are discussed in Sec.~\ref{seccavityeffect}. Section~\ref{secrotation} describes the physics of the rotation of the noise ellipse and traces parallels with the two well known techniques, homodyne detection and Pound-Drever-Hall whenever possible. Section~\ref{secclassroom} discusses a simple experiment for demonstrating the effect. Concluding remarks are presented in Sec.~\ref{secconclusions}. 

\section{QUANTUM NOISE AND FIELD QUADRATURES}
\label{seclightnoise}

In the classical description of light a complex number is used to represent the oscillating electric field. However, when microscopic features are considered, a precise description must include its quantum features: light travels in small packets of minimum energy, the photons. The measurement of light intensity is usually performed by employing the photoelectric effect in which one photon is converted into one free electron in the detector. Because of this grainy character of light, the photocurrent cannot be described in the ideal case by a continuous function, but rather as a sequence of steps. In practice the flux of photons in the laser beam is so huge that no photodetector working in this regime can distinguish individual steps in the photocurrent, and a continuous function is obtained. Nevertheless, the statistics of photon arrival times is transferred to the photocurrent. The statistics is called ``shot noise'' if the times are random, resulting in Poisson statistics. The intensity fluctuations of an ideal laser operating far above threshold are Poisson. A more regular photon flux would result in less photocurrent noise. This sub-Poissonian light is called ``squeezed light,''\cite{squeezeprimer_ajp88,ekertknightsqueeze_ajp89} and cannot be obtained solely by common (linear) optical transformations such as lenses and beam splitters. Squeezing is a quantum property, intimately related to entanglement.\cite{reviewbraunsteinvanloock_rmp05,qucakes_ajp00}

\begin{figure}[ht]
\centering
\epsfig{file=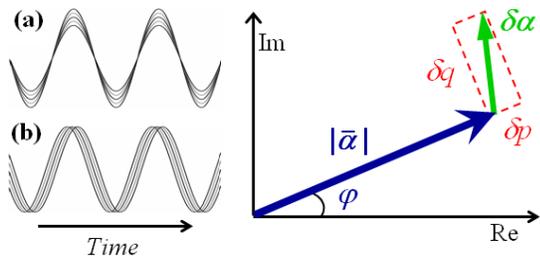,scale=0.5}
\caption{Representation of the light field in the complex plane and its fluctuations. The vector is the complex field mean amplitude $\bar\A=|\bar\A|\exp(i\varphi)$. The decomposition of the fluctuation $\de\A$ in the direction of the mean value and in the orthogonal one defines the amplitude $p$ and phase $q$ quadratures, respectively. In the time domain a field presenting the amplitude fluctuations would resemble the inset (a), and the phase fluctuations would correspond to (b).}
\label{quadamplifase}
\end{figure}

For a laser-like field, with a typical mean photon flux on the order of $10^{16}$\,s$^{-1}$, intensity measurements result in well defined macroscopic numbers $I(t)$ with very small fluctuations $\de I(t)$ in time. In this case the quantum fluctuations may be treated as classical fluctuations, although their physical origin cannot be explained classically (the semi-classical approach\cite{fabresemiclassical_leshouches92}), and we may once more represent the electric field of light by a complex number. However, to account for the uncertainty principle, this complex amplitude is no longer perfectly defined and fluctuates in time,
\be
\label{adet}
\A(t)=\bar\A+\de\A(t),
\ee
where 
\be
\ave{\A(t)}\equiv \bar\A=|\bar\A|\,\exp(i\varphi)
\ee
is its mean value and $\de\A(t)$ its fluctuations. A bright beam means that $|\de\A(t)|\ll|\bar\A|$. As usual, the field intensity (photons per time interval) is $I(t)=\A^*(t)\A(t)$. We use Eq.~(\ref{adet}) to obtain for the intensity fluctuations,
\be
\label{defdeit}
\de I(t)=\bar\A^*\de\A(t)+ \bar\A\de\A^*(t),
\ee
where the second-order term has been neglected assuming a bright beam. We see from Eq.~(\ref{defdeit}) and Fig.~\ref{quadamplifase} that the intensity fluctuations are proportional to the projection of $\de\A(t)$ onto the direction of $\bar\A$ in the complex plane. To avoid a dependence on the absolute value of the field we normalize this projection by the latter, resulting in the definition of the first important physical quantity we will study, the {\it amplitude quadrature} fluctuation,
\be
\de p(t)=\e^{-i\varphi}\de\A(t)+ \e^{i\varphi}\de\A^*(t),
\ee
so that $\de I(t)=|\bar\A|\de p(t)$. Because $\de p(t)$ is proportional to the intensity, it is directly measured by photodetectors; $\de p$ is related to the fluctuations of the complex field absolute amplitude of Fig.~\ref{quadamplifase}. 

The second important quantity associated with the electric field of light is the phase. According to quantum mechanics, the phase is the conjugate observable of intensity for bright light fields.\cite{fabreILOQ} 
In Fig.~\ref{quadamplifase} a small phase change of $\bar\A$ corresponds to a fluctuation in the direction orthogonal to $\de p(t)$, and is therefore called the {\it phase quadrature} fluctuation,
\be
\de q(t)=-i[\e^{-i\varphi}\de\A(t)- \e^{i\varphi}\de\A^*(t)].
\ee
It is the projection of $\de\A(t)$ in the direction orthogonal to $\bar\A$. Phase information can be only accessed indirectly by converting it to amplitude information by means of an interference effect.

The time dependence assumed for the quadrature fluctuations implicitly assumes a multimode description of the light field, that is, several frequency components. It is natural to consider these ideal well defined frequency complex amplitudes,
\be
\label{defanu}
\de\A(\nu)=\!\int^{T/2}_{-T/2}\e^{i2\pi\nu\,t}\de\A(t)dt, \qquad [\de\A(\nu)]^*=\de\A^*(-\nu),
\ee
which give us better physical insight into the fluctuations. The integration time $T$ is much longer than the typical fluctuation time scale of the system, and we therefore take $T\rightarrow\infty$ to simplify the calculations. This Fourier transform is evaluated around the optical frequency $\nu_0\sim10^{14}$\,Hz ($\nu\ll\nu_0$), because photodetectors cannot resolve the fast oscillations of an optical field and record only its average intensity over many optical cycles. The mean field $\bar\A$ is thus represented by the zero frequency component, $\nu=0$, which is called the carrier. Practically all the energy present in the light field belongs to the carrier. Its quantum character is not considered here because of technical limitations in distinguishing photon numbers at this intensity level, although they may show very interesting quantum features.\cite{schleich_nature87} Therefore, it is justified to consider it as a classical field for our purposes.

The frequency components around the optical carrier are called sidebands. The analysis frequency $\nu$ in which to observe them is chosen according to the experiment (typically radio frequencies, $\nu\sim1$\,MHz to $100$\,MHz). In the time domain they are related to the intensity and phase modulations of the carrier.\cite{bachorGEQO} Therefore, in the frequency domain the quantum states to be considered belong to the sideband region. If no energy exists in these modes, they are in the vacuum state, which also possesses intensity and phase uncertainties but has zero mean values. Thus, light with sidebands in the vacuum state presents amplitude and phase noise -- another way of understanding the shot noise. In contrast, when light is externally modulated (such as in the Pound-Drever-Hall method), we may associate this energy with the existence of photons in the sidebands. Similarly, if we apply a specific quantum dynamics to the beam, the noise in one quadrature becomes smaller than the shot noise at the expense of increasing the conjugate noise. This situation represents a squeezed state of light, and in this case the sidebands are also populated with photons. For typical experimentally observed squeezing levels, the mean number of photons per frequency interval is on the order of unity.\cite{wallsmilburnQO}

The single-frequency quadrature components are given by
\ba
\label{pqomega1}
\de p(\nu)&=&\e^{-i\varphi}\de\A(\nu)+ \e^{i\varphi}\de\A^*(-\nu),\\
\label{pqomega2}
\de q(\nu)&=&-i[\e^{-i\varphi}\de\A(\nu)- \e^{i\varphi}\de\A^*(-\nu)].
\ea
They have the form of beats between the carrier optical frequency and the sidebands symmetrically located at the frequencies $\pm\nu$ around it (normalized by the carrier absolute amplitude). Thus, despite the fact that the sidebands have a very low photon number, their effect can be recorded by normal photodetectors because of the enormous carrier power. 

Amplitude and phase fluctuations differ only by a phase: the amplitude quadrature is in phase with the mean field complex amplitude, and the phase quadrature is in quadrature with it. It is possible to convert phase to amplitude fluctuations by manipulating the relative phase between the carrier and sidebands. To show this conversion we consider how the amplitude quadrature of Eq.~(\ref{pqomega1}) would change if we could shift the carrier or the sidebands by a phase $\theta$ tunable by some physical means. By adding this phase, for example, to the complex sideband $\de\A(\nu)$, the amplitude quadrature goes to
\begin{subequations}
\label{xtheta1}
\ba
\lefteqn{\de p(\nu) \rightarrow  \de p'(\nu) = \e^{i\theta}\e^{-i\varphi}\de\A(\nu)+ \e^{i\varphi}\de\A^*(-\nu)}\; \\
&=& \e^{i\theta/2}[\e^{-i(\varphi-\theta/2)}\de\A(\nu)+ \e^{i(\varphi-\theta/2)}\de\A^*(-\nu)].
\ea
\end{subequations}
By varying $\theta$ from $0$ to $2\pi$, the amplitude quadrature assumes the value of the phase quadrature for one possible $\theta$, $\theta=\pi$ (the leading phase plays no role). A similar effect occurs by varying the other sideband phase. The carrier phase could be varied as well, $\varphi\rightarrow\theta+\varphi$. In this case, $\de p$ changes according to
\be
\label{xtheta2}
\de p(\nu)\rightarrow \de p'(\nu)=\e^{-i(\varphi+\theta)}\de\A(\nu)+ \e^{i(\varphi+\theta)}\de\A^*(-\nu),
\ee
which is equal to the phase quadrature for $\theta$ equal to $\pi/2$ and $3\pi/2$. Homodyne detection is described by an identical expression, with the difference that $\theta$ represents the local oscillator phase. Therefore, the possibility of independently varying the phases of the carrier and sidebands allows complete conversion between the quadratures. This effect is what an optical cavity realizes close to resonance.

\section{OPTICAL CAVITY}
\label{secoticalcavity}

An optical cavity is a region of space delimited by mirrors where light is confined for some time. Only certain frequencies of light which fulfill a resonance condition are able to probe this region. Every resonance has a dispersive character. Thus, the different frequency components of a light beam reflected by an optical cavity close to resonance experience different phase shifts.

\begin{figure}[ht]
\centering
\epsfig{file=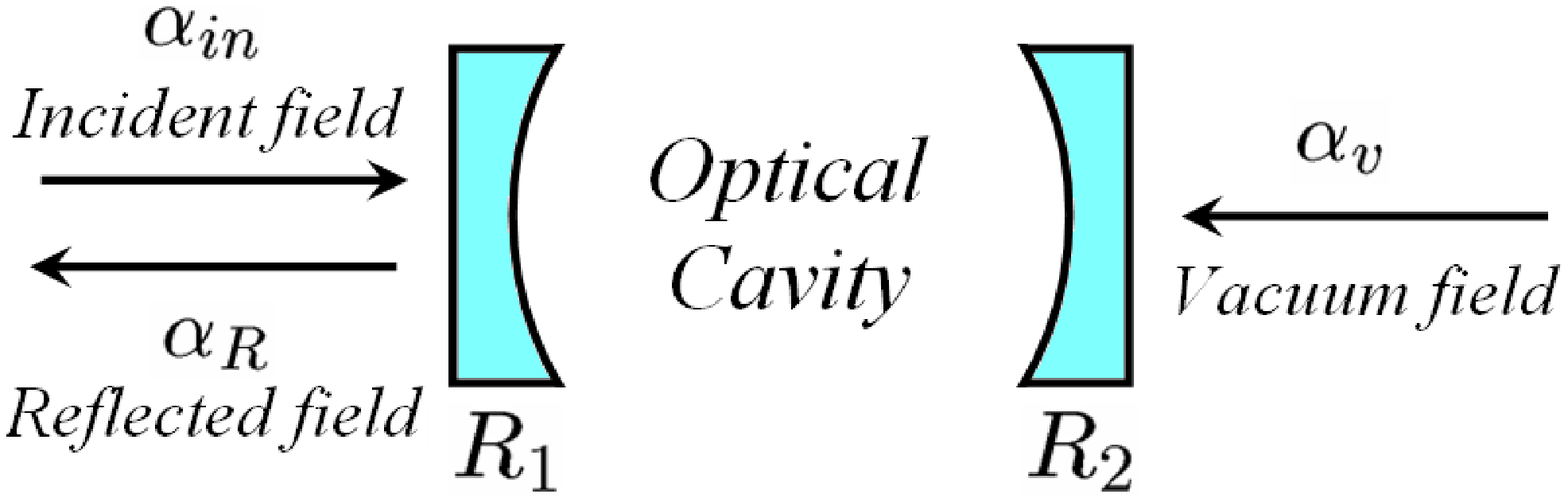,scale=0.28}
\caption{Linear optical cavity with coupling mirror showing the reflectivity $R_1$ and the output mirror with reflectivity $R_2$. The reflected field amplitude $\A_R$ is the sum of the incident amplitude $\A_{\rm in}$ with the vacuum $\A_v$ coupled by the output mirror and losses.}
\label{cavotica}
\end{figure}

Consider the situation depicted in Fig.~\ref{cavotica}. The optical cavity has a coupling mirror with intensity reflection $R_1$ and transmission $T_1$, and an output mirror, with reflection $R_2$ and transmission $T_2\ll1$ representing spurious losses. The amplitude reflection and transmission coefficients are, respectively, $r_j^2=R_j$ and $t_j^2=T_j$, $j=1,\,2$. The cavity resonance frequency closest to the incident light beam is denoted by $\nu_c$. Three important parameters characterize the optical cavity: the finesse, $F=\pi\,\frac{(R_1\,R_2)^{1/4}}{1-\sqrt{R_1\,R_2}}$, the free spectral range $\D\nu_c=c/L$, (where $L$ is the cavity perimeter, that is, the distance traveled by light in one round trip\cite{corresponds}), and the resonance bandwidth (or full width at half maximum), $\de\nu_c=\D\nu_c/F$. They represent the amplification of the field inside the cavity compared to the incident field, the inverse of the round-trip time of a photon, and the inverse of the average time a photon remains inside the cavity, respectively. 

\begin{figure}[ht]
\centering
\epsfig{file=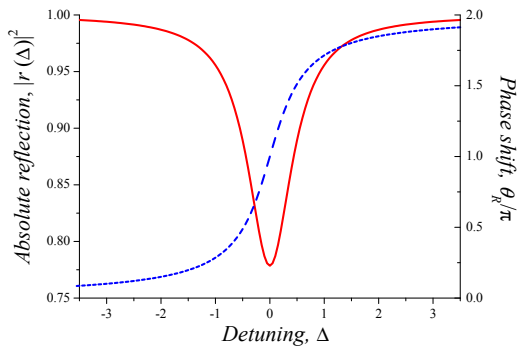,scale=0.25}
\caption{Squared modulus (continuous line) and phase (dashed line) of $r(\D)$ as functions of the carrier-cavity detuning parameter $\D$ relative to the cavity bandwidth. The numerical values $R_1=95.0\%$ and $R_2=0.3\%$ were used.}
\label{romega}
\end{figure}

An optical field $\A_{\rm in}(t)=\bar\A_{\rm in}+\de\A_{\rm in}(t)$ incident on the cavity coupling mirror generates a reflected beam $\A_R=\bar\A_R+\de\A_R(t)$ after mixing with the vacuum fluctuations $\A_v(t)=\de\A_v(t)$ that couple to the cavity through the output mirror and spurious losses. If we define the analysis frequency relative to the cavity bandwidth, 
\be
\nu'=\nu/\de\nu_c,
\ee
they are related to each other by the expression\cite{fabreILOQ}
\be
\label{aRain}
\A_R(\nu')=r(\D+\nu')\,\A_{\rm in}(\nu')+t(\D+\nu')\A_v(\nu'),
\ee
where the optical cavity amplitude reflection $r(\D)$ and transmission $t(\D)$ functions are
\begin{subequations}
\ba
r(\D)&=&\frac{r_1-r_2\exp(i2\pi\D/F)}{1-r_1r_2\exp(i2\pi\D/F)},\\
t(\D)&=&\frac{t_1\,t_2\exp(i\pi\D/F)}{1-r_1r_2\exp(i2\pi\D/F)},
\ea
\end{subequations}
and 
\be
\D=(\nu_0-\nu_c)/\de\nu_c
\ee
is the detuning, relative to the cavity bandwidth, between the carrier frequency $\nu_0$ and the closest cavity resonance frequency $\nu_c$. These expressions are easily deduced by considering the sum of all reflected waves inside the cavity.

The squared modulus and phase $\theta_R$ of $r(\D)$ are plotted in Fig.~\ref{romega} as a function of $\D$. The lorentzian curve $|r(\D)|^2$ represents an attenuation in the reflected beam relative to the incident one. The phase $\theta_R$ varies from $0$ to $2\pi$ across resonance such that the relative phase between off-resonant and exactly resonant fields is $\pi$. By energy conservation, $|r(\nu)|^2+|t(\nu)|^2=1$. The phase of $t(\D)$ has the same shape, but varies from $0$ to $\pi$ across the resonance, similarly to simple mechanical resonances. 

\section{CAVITY EFFECT ON THE FIELD QUADRATURES}
\label{seccavityeffect}

\begin{figure}[ht]
\centering
\epsfig{file=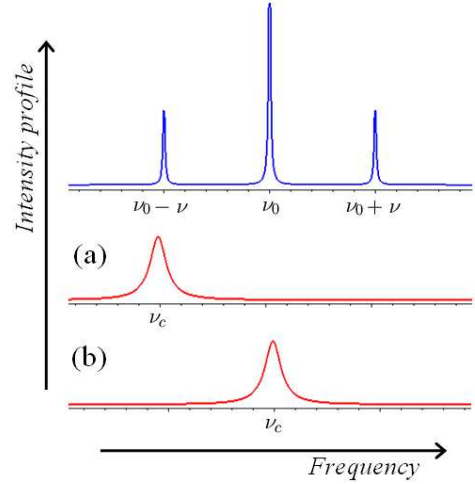,scale=0.5}
\caption{Representation of the light field frequency components, carrier and sidebands (upper curve), and cavity resonance transmission profile (lower curve). The cavity is either resonant with (a) one sideband or (b) the carrier.}
\label{cavressonante}
\end{figure}

How does an empty optical cavity realize the effects given by Eqs.~(\ref{xtheta1}) and (\ref{xtheta2})? To understand the answer we recall the multimode description of bright fields in which each frequency mode is given by Eqs.~(\ref{pqomega1}) and (\ref{pqomega2}). The frequency components around the carrier are recorded by the photodetector, and the resulting beatnote signal at a chosen frequency defines the amplitude fluctuations. Photodetectors are in principle insensitive to the phase fluctuations of light, but if the light beam is previously reflected by an optical cavity, additional phase shifts appear between the carrier and sidebands as a consequence of the cavity dispersion (see Fig.~\ref{cavressonante}). The final effect is that the reflected field amplitude fluctuations can provide information on the incident field phase fluctuations. The exact noise conversion dependence on the sideband frequency and cavity detuning is deduced in the following. 

We apply Eq.~(\ref{aRain}) to the field mean amplitude to determine how the carrier is affected by the reflection,
\be
\label{carrierout}
\bar \A_R=r(\D)\,\bar \A_{\rm in},
\ee
where a zero mean amplitude has been used for the vacuum field. We choose $\varphi_{\rm in}=0$, that is, the incident carrier field is chosen as the phase reference. The same equation applied to the sidebands results in
\begin{subequations}
\begin{align}
\label{sidebandsout}
\de\A_R(\nu')& = r(\D+\nu')\,\de\A_{\rm in}(\nu') +t(\D+\nu')\,\de\A_v(\nu'),\\
\de\A^*_R(-\nu')& = r^*(\D-\nu')\,\de\A^*_{\rm in}(-\nu') +t^*(\D-\nu')\,\de\A^*_v(-\nu').
\end{align}
\end{subequations}
In this way the reflected field frequency component that is resonant to the cavity undergoes the attenuation $|r(\D)|$ and the phase delay $\theta_R(\D)$ depicted in Fig.~\ref{romega},
\be
\label{expthetar}
\exp[i\theta_R(\D)]= r(\D)/|r(\D)|.
\ee
It is also contaminated by the vacuum field through the coefficient $|t(\D)|$.

Once we know the cavity effect on the field carrier and sidebands, we can deduce their composite effect on the reflected beam amplitude quadrature $\de p_R$. To show its explicit dependence on carrier and sidebands, it may be written as
\be
\label{pomegaout}
\de p_R(\nu\,')=\frac{\bar\A^*_{R}}{|\bar\A_R|}\,\de\A_R(\nu\,') +\frac{\bar\A_R}{|\bar\A_R|}\,\de\A_R^*(-\nu\,').
\ee
We substitute in Eq.~\eqref{pomegaout} the results of Eqs.~(\ref{carrierout})--(\ref{expthetar}) and obtain
\ba
\label{pomegaoutin}
\de p_R(\nu\,')&=&\mathrm{e}^{-i\theta_R(\D)}\,r(\D+\nu\,')\,\de\A_{\rm in}(\nu\,')  \nonumber\\
&&{}+\mathrm{e}^{i\theta_R(\D)}\,r^*(\D-\nu\,')\,\de\A_{\rm in}^*(-\nu\,') \nonumber\\
&&{}+\mathrm{e}^{-i\theta_R(\D)}\,t(\D+\nu\,')\,\de\A_{v}(\nu\,')  \nonumber\\
&&{}+\mathrm{e}^{i\theta_R(\D)}\,t^*(\D-\nu\,')\,\de\A_{v}^*(-\nu\,').
\ea

Equation~\eqref{pomegaoutin} should be compared to Eqs.~(\ref{xtheta1}) and~(\ref{xtheta2}). For simplicity, the output mirror transmission is set to zero in the following discussion ($|r(\D)|=1$ and $|t(\D)|=0$), so that the last two terms of Eq.~(\ref{pomegaoutin}) are zero. It is also supposed that the cavity acts only on one frequency component at a time ($\nu\,'\gg1$), as depicted in Fig.~\ref{cavressonante}. Only one of the phase factors appearing in Eq.~(\ref{pomegaoutin}) is different from unity in this situation. When one sideband is close to the cavity resonance (see Fig.~\ref{cavressonante}(a)), Eq.~(\ref{pomegaoutin}) takes the form
\be
\label{prnu}
\de p_R(\nu\,')=\e^{i\theta_R(\D+\nu\,')}\,\de\A_{\rm in}(\nu\,') +\de\A_{\rm in}^*(-\nu\,'),
\ee
which is analogous to Eq.~(\ref{xtheta1}). If the cavity is resonant to the carrier (see Fig.~\ref{cavressonante}(b)), then
\be
\label{prnucarrier}
\de p_R(\nu\,')=\e^{-i\theta_R(\D)}\,\de\A_{\rm in}(\nu\,') +\e^{i\theta_R(\D)}\,\de\A_{\rm in}^*(-\nu\,'),
\ee
as in Eq.~(\ref{xtheta2}). In this case the carrier plays the same role as the local oscillator in homodyne detection, but without introducing extra noise in the beam. In conclusion, the hypothetical controllable phase $\theta$ considered in the discussion of Sec.~\ref{seclightnoise} is precisely $\theta_R(\D)$. 

In the general case when the simplification \mbox{$\nu\,'\gg1$} does not apply, the dephasings of the three frequency components interfere. By decreasing the analysis frequency even further, the limiting situation where the carrier and sidebands cannot be distinguished from each other inside the cavity bandwidth is reached ($\nu\,'\ll1$): because all components acquire the same phase shift, there is no quadrature conversion. Thus, there is a minimum value of $\nu\,'$ for which the phase noise is completely converted to amplitude noise. This value can be shown to be $\nu\,'=\sqrt{2}$ by imposing $\theta_R(\D)-\theta_R(\D-\nu\,')=\pi/2$ in Eq.~(\ref{pomegaoutin}). If $\nu\,'<\sqrt{2}$, quadrature conversion is partial; if $\nu\,'\geq\sqrt{2}$, it is complete.

The same frequency-dependent dephasing is exploited by the Pound-Drever-Hall technique, but in the opposite sense: classical sidebands resulting from external phase modulation serve as references to the mean field. Even a small detuning between the laser and the cavity converts the phase to a detectable amplitude modulation, such that the laser intensity fluctuation works as a sensitive probe for the detuning.

\begin{figure}[ht]
\centering
\epsfig{file=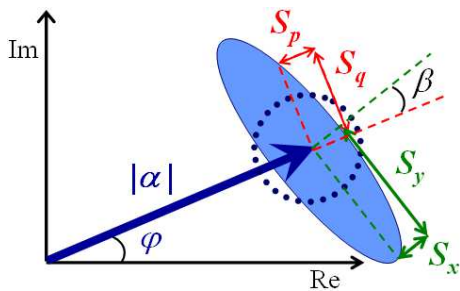,scale=0.5}
\caption{Noise ellipse representation in the complex plane (shaded ellipse). The dotted circle represents the shot noise. The ellipse size compared to the mean value is exaggerated for ease of visualization.}
\label{noiseelipse}
\end{figure}

The final expression for $\de p_R$ as a function of the input quadratures $\de p_{\rm in}$ and $\de q_{\rm in}$ is obtained by inverting Eqs.~(\ref{pqomega1}) and (\ref{pqomega2}) for the incident field and substituting the results in Eq.~(\ref{pomegaoutin}),
\ba
\label{pr}
\de p_R(\D,\nu\,') &=& g_p\,\de p_{\rm in}(\nu\,') +ig_q\,\de q_{\rm in}(\nu\,')  \nonumber\\
&&{}+g_{vp}\,\de v_p(\nu\,') + i\,g_{vq}\,\de v_q(\nu\,'),
\ea
where $\de v=\mathrm{e}^{i\varphi}(\de v_p+i\de v_q)/2$ is the vacuum fluctuation, and
\begin{subequations}
\begin{align}
2g_p&= \e^{-i\theta_R(\D)}r(\D+\nu\,') +\e^{i\theta_R(\D)}r^*(\D-\nu\,'),\\
2g_q&= \e^{-i\theta_R(\D)}r(\D+\nu\,') -\e^{i\theta_R(\D)}r^*(\D-\nu\,'),\\
2g_{vp}&= \e^{-i\theta_R(\D)}t(\D+\nu\,') +\e^{i\theta_R(\D)}t^*(\D-\nu\,'),\\
2g_{vq}&= \e^{-i\theta_R(\D)}t(\D+\nu\,') -\e^{i\theta_R(\D)}t^*(\D-\nu\,'),
\end{align}
\end{subequations}
are functions dependent on the cavity characteristics and detuning.

The coefficients $g_p$ and $g_q$ involve the dependence of $\de p_R(\nu\,')$ on the input field quadratures, and $g_{vp}$ and $g_{vq}$ are related to the vacuum contributions. The effect of these spurious losses is to remove photons from the resonant field component, partially taking it into the vacuum: if one sideband is resonant, the reflected beam noise tends to shot noise; in the case of the carrier, only the ``local oscillator'' is attenuated, and no loss of noise information occurs because the sidebands are totally reflected. If no cavity losses are present, then $|g_{vp}|=|g_{vq}|=0$ and $|g_p|^2+|g_q|^2=1$, and we may write $|g_p|=\cos\Theta$ and $|g_q|=\sin\Theta$ (where $\Theta$ is a cavity-dependent parameter) to show the analogy with homodyne detection. The main difference between these two techniques is the functional form of $\Theta$ on the parameters. Although it varies linearly with the relative phase between the local oscillator and the measured field in homodyne detection, in the case treated here $\Theta$ results from the interference of three phase shifts, each of which varies as the arctangent of the cavity detuning respective to its frequency component. 

\section{ROTATION OF THE NOISE ELLIPSE}
\label{secrotation}

The noise at the analysis frequency $\nu$ is defined as the power present in the fluctuations of this frequency component. It is measured relative to the shot noise, defined here as the quadrature noise of a light beam with sidebands in the vacuum state. The noise spectrum $S_{X_\phi}(\nu)$ of a generalized quadrature $\de X_\phi=\exp(-i\phi)\A+\exp(i\phi)\A^*$ is calculated by the Wiener-Khintchine theorem,\cite{mandelwolfOCQO} which results in the relation
\be
\label{wienerkhintchine}
S_{X_{\phi}}(\nu)\de(\nu-\nu\,'')=\ave{\de X_{\phi}(\nu)\de X_{\phi}(-\nu\,'')}.
\ee
This ideal noise spectrum is a very good approximation to the actual measured quantity if the measurement integration time is much longer than the typical time scale of the system variations, as assumed in Eq.~(\ref{defanu}). The delta function in Eq.~\eqref{wienerkhintchine} means that the treatment assumes perfectly defined frequencies; it disappears as soon as the finite precision of the frequency definition (bandwidth) is included.

The noise ellipse is a representation of the uncertainty statistics of the carrier intensity and phase at a certain analysis frequency (see Fig.~\ref{noiseelipse}). It can be rigorously defined as a contour of the Wigner function describing the field in phase space.\cite{gardinerQN} For a shot-noise limited beam it is a circle with unit radius. For a squeezed state, its minor axis represents the maximally squeezed quadrature, and its major axis stands for the anti-squeezed one. If either of these axes is aligned with the mean field complex amplitude, then the optima squeezed and anti-squeezed quadratures correspond to the amplitude and phase quadratures. The optima noise powers are denoted as $S_x$ and $S_y$, and $S_p$ and $S_q$ the amplitude and phase quadratures noise, respectively. The uncertainty principle imposes $S_x S_y\geq1$ and $S_p S_q\geq1$. If the ellipse axes are not aligned with $\bar\A$, there is a correlation between $\de p$ and $\de q$.

\begin{figure}[ht]
\centering
\epsfig{file=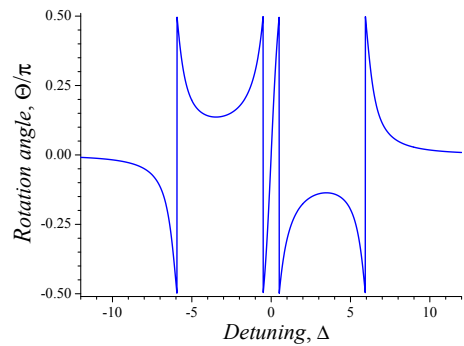,scale=0.25}
\caption{The rotation angle $\Theta$ between the field carrier and the noise ellipse introduced by the cavity as a function of $\Delta$ for $\nu\,'=6$.}
\label{rotangle}
\end{figure}

The angle between the mean amplitude and the ellipse axis associated with $S_x$ is $\beta$. An optical cavity close to resonance with the optical field, by realizing the effects in Eqs.~(\ref{xtheta1}) and (\ref{xtheta2}), rotates this relative angle. In fact, the quantity $\Theta$ is precisely the angle added by the cavity, presented as a function of detuning in Fig.~\ref{rotangle}. Its shape resembles the Pound-Drever-Hall error signal\cite{pdh_ajp01} due to the fact that the latter is chosen as the dispersive part of the cavity action over the sidebands,\cite{levenson_applphysb83} which is directly related to $\Theta$.

\begin{figure}[ht]
\centering
\epsfig{file=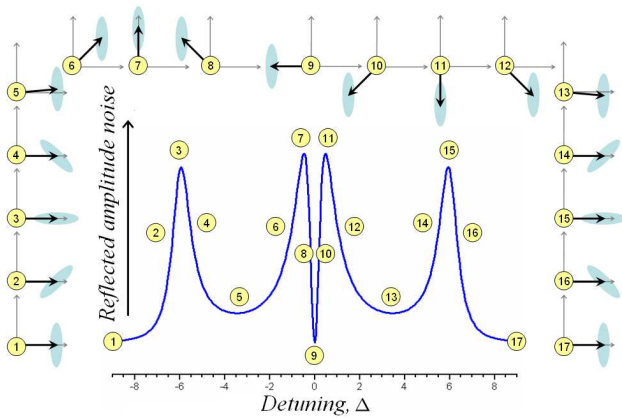,scale=0.34}
\caption{Phase rotation of the noise ellipse as a function of the carrier-cavity detuning parameter $\D$ relative to the cavity bandwidth. The central curve is $S_R(\D,\nu\,')$ [Eq.~(\ref{sr})], and the frames around it represent, for the corresponding numbered detuning, the reflected field in the complex plane (see Fig.~\ref{noiseelipse}). For $\nu\,'=6$, $S_p<S_q$, $R_1=95.0\%$, and $R_2=0.3\%$.}
\label{explicarotelipse}
\end{figure}

The final step in our derivation is to obtain the amplitude noise spectrum $S_R(\D,\nu\,')$ of the reflected field, calculated from Eq.~(\ref{wienerkhintchine}) and applied to Eq.~(\ref{pr}),
\be
\label{sr}
S_R(\D,\nu\,')=|g_p|^2\,S_p(\nu\,')+|g_q|^2\,S_q(\nu\,')
+|g_{vp}|^2+|g_{vq}|^2,
\ee
where $S_{vp}(\nu)=S_{vq}(\nu)=1$ is by convention the vacuum noise and $\ave{\de v_{p,q}(\nu)\de v_{p,q}(-\nu\,'')}=\de(\nu-\nu\,'')$ has been used for the vacuum field. The noise ellipse was assumed to be aligned with the carrier ($\beta=0$).

Figure~\ref{explicarotelipse} presents a typical noise power curve as a function of the cavity detuning parameter $\D$ for $\nu\,'=6$. We chose $S_p>S_q$ in this example, such that the noise profile contains four peaks (four valleys would appear instead if $S_q>S_p$). Around the curve, small frames show the reflected field representation in the complex plane, each corresponding to a labeled detuning. For large detuning (frame~1), the cavity has no effect on the beam, and amplitude noise is observed. As the first sideband is brought close to resonance, with $\D\approx-\nu\,'$ (frames~2--4), it undergoes a phase shift, causing a rotation in the noise ellipse while the mean field remains undisturbed. At exact sideband resonance ($\D=-\nu\,'$, frame~3), the ellipse rotates by $\pi/2$, and phase noise is observed. Only one phase to amplitude noise conversion occurs when the sideband passes through the resonance [Eq.~(\ref{xtheta1})]. A small fraction of vacuum fluctuations originating from the spurious losses contaminates the quantum sideband, making the noise ellipse a bit closer to the shape of a circle. For $|\D|\lesssim1$, the carrier is resonant, while the sidebands are far off-resonance (frames~6--12). The carrier experiences a $\pm\pi/2$ phase shift at $\D=\mp0.5$, giving rise to two complete conversions [Eq.~(\ref{xtheta2})] between quadratures (frames~7 and~11). The vacuum now contaminates only the carrier and attenuates it. Because only the local oscillator is attenuated, there is no vacuum contribution to the noise measurement, and perfect phase noise is observed. Hence, peaks 7 and 11 are slightly higher than peaks 3 and 15. The carrier rotates by $\pi$ at exact resonance (frame 9), without any affect on the reflected beam noise spectrum, and we observe amplitude noise once more. The noise ellipse rotates again as the other sideband enters resonance (frames 14--16). This noise profile gives complete information about the noise ellipse, allowing for the reconstruction of the sideband state\cite{noise} by quantum tomography.\cite{pengcavtomo_josab00,giacomellicavtomo_pra02}

All three frequencies dephasings interfere as the analysis frequency is decreased. Peaks 3 and 7 (as well as peaks 11 and 15) approach each other and finally collapse in one single peak for $\nu\,'=\sqrt{2}$. Below this analysis frequency their amplitudes decrease, because the carrier and sidebands cannot be perfectly distinguished inside the cavity bandwidth, and a partial ellipse rotation occurs.

\begin{figure}[ht]
\centering
\epsfig{file=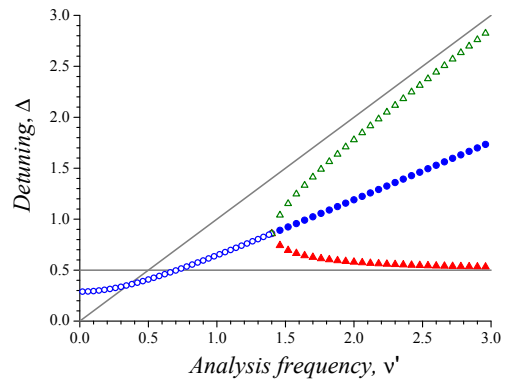,scale=0.27}
\caption{Detunings $\D>0$ for which $S_R(\D,\nu\,')$ has zero derivatives as a function of the analysis frequency $\nu\,'$. Each symbol represents a different zero derivative point. At $\nu\,'=\sqrt{2}$ a single detuning with zero derivative (open circles) gives rise to three such points as $\nu\,'$ increases, because the complete conversion of phase to amplitude noise begins to be possible (triangles). The gray curves help to visualize the asymptotic behavior. Losses are assumed to be zero ($R_\mathrm{min}=1$).}
\label{zerosrotelipse}
\end{figure}

These regimes can be better characterized by observing the detunings for which $S_R(\D,\nu\,')$ has zero derivatives as $\nu\,'$ is varied (see Fig.~\ref{zerosrotelipse}). For symmetry reasons only $\D>0$ is considered, and a perfect cavity is now assumed ($R_2=0$). For $\nu\,'<\sqrt{2}$, there is only one detuning where a zero derivative occurs, because phase to amplitude conversion is partial (open circles). As $\nu\,'$ increases above $\sqrt{2}$, three positive detunings with zero derivatives appear in total. Two of them correspond to the full conversion of phase to amplitude noise (triangles), and the third corresponds to the inflection point between the two complete noise conversions (full circles). As $\nu\,'\rightarrow\infty$, the sidebands and carrier phase shifts no longer interfere. The first full conversion, due to the carrier rotation, occurs at $\D\approx0.5$ (full triangles), and the second one, due to the sideband rotation, occurs at $\D\approx\nu\,'$ (open triangles), as expected. The gray lines show this asymptotic behavior.

It can be shown that the zero derivative points also depend on the losses. As $R_2$ is varied for a given analysis frequency, the smallest detuning for which complete conversion occurs goes to zero as $R_2\rightarrow R_1$. For this reason the Pound-Drever-Hall method has a very steep error signal close to exact resonance and is most effective when $R_1=R_2$.

\section{Classroom experiment}
\label{secclassroom}

\begin{figure}[ht]
\centering
\epsfig{file=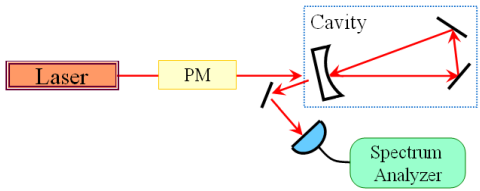,scale=0.4}
\caption{Schematic of the experimental setup required to perform a classroom demonstration of the effect. PM: phase modulator.}
\label{setup}
\end{figure}

A classroom demonstration of this effect can be performed with the same apparatus used in the Pound-Drever-Hall technique:\cite{pdh_ajp01} an optical cavity, a laser, a phase modulator, a photodetector, and a spectrum analyzer (Fig.~\ref{setup}).

The cavity should be stable, possibly built as a rigid metallic body. The ring geometry would be more convenient than the linear one, because it does not require an optical isolator, wave plate, and polarizing beam splitter. Because the quantum noise is very weak, it would be more practical to modulate the laser phase (to create strong sidebands). The cavity bandwidth has to be compatible with this frequency (see Figs.~\ref{explicarotelipse} and \ref{zerosrotelipse}). The beam reflected by the cavity is measured with a photodetector. The fast components of the photocurrent are then sent to a spectrum analyzer set to measure at a single frequency, the same used to modulate the beam phase. As the laser frequency is linearly scanned around the cavity resonance (the cavity transmission or reflection profile can be monitored to check for the resonance condition), the noise profile should show the features in Fig.~\ref{explicarotelipse}. It would be interesting to change the laser modulation frequency and check how it affects the noise profile (Fig.~\ref{zerosrotelipse}). Alternatively, we could employ a diode laser, which possesses strong phase noise but low amplitude noise, to eliminate the phase modulator. In this case we would measure directly the laser phase noise; however, this measurement might require us to amplify the photocurrent signal. It is natural that such a demonstration be part of a lecture on the optical cavity and on the quantum noise of light.

\section{CONCLUSION}
\label{secconclusions}

A detailed description of the physical process that allows an empty optical cavity to convert phase to amplitude noise of a bright light beam has been presented. We have considered amplitude and phase fluctuations as frequency sidebands around the carrier optical frequency. For measurements based on photodetectors, only the light fluctuations in phase with the carrier (the amplitude quadrature) can be measured. However, the dispersive character of an optical cavity resonance can be used to introduce a relative phase between the carrier and the sidebands. For certain detunings the conversion of phase to amplitude noise in the beam reflected by the cavity occurs: in phase space, the field carrier and noise ellipse rotate relative to each other. The carrier has a completely analogous role to the local oscillator used in the homodyne detection technique, with the advantage of perfect spatial and temporal overlap and no additional noise. Furthermore, every quadrature is easily accessible, as in homodyne detection, giving complete information about the field quantum state. The rotation of the noise ellipse of light can be viewed as an application of the Pound-Drever-Hall technique in the opposite sense: instead of using classical sidebands to measure the detuning between carrier and a reference cavity, it employs the carrier as a local oscillator to the quantum sidebands of bright light beams.

\begin{acknowledgements}

I gratefully thank Paulo Nussenzveig and Marcelo Martinelli for their advice throughout my student years, and for posing to me the questions discussed in this article: thinking about their intuitive meaning has provided me a lot of the joy of physics. I dedicate this work to them. I also thank Katiuscia N.\ Cassemiro for valuable suggestions, Wolfgang Schleich for his encouragement, and Gerd Leuchs for his interest in this manuscript. This work was supported by Funda\c{c}\~ao de Amparo \`a Pesquisa do Estado de S\~ao Paulo (FAPESP) and the European Commission through the SCALA network.
\end{acknowledgements}


\begin{thebibliography}{99}

\bibitem{nielsenchuangQCQI} 
M. A. Nielsen and I. L. Chuang, {\it Quantum Computation and Quantum Information} (Cambridge University Press, Cambridge, 2000).

\bibitem{reviewbraunsteinvanloock_rmp05}
S. L. Braunstein and P. van Loock,
``Quantum information with continuous variables,''
Rev. Mod. Phys. {\bf 77}, 513--577 (2005).

\bibitem{fabreILOQ}
G. Grynberg, A. Aspect, and C. Fabre, {\it An Introduction to Lasers and Quantum Optics} (Cambridge University Press, Cambridge, 2001).

\bibitem{epr_physrev35}
A. Einstein, B. Podolsky, and N. Rosen,
``Can quantum-mechanical description of physical reality be considered complete?,''
Phys. Rev. {\bf 47}, 777--780 (1935).

\bibitem{jozsa_procrsoca03}
R. Jozsa and N. Linden,
``On the role of entanglement in quantum-computational speed-up,''
Proc. R. Soc. A {\bf 459}, 2011--2032 (2003).

\bibitem{forcer_quantuminformcompu03}
T. M. Forcer, A. J. G. Hey, D. A. Ross, and P. G. R. Smith,
``Superposition, entanglement and quantum computation,''
Quantum Inform. Compu. {\bf 2}, 97--116 (2002).

\bibitem{machzehnder_ajp95}
P. Nachman,
``Mach-Zehnder interferometer as an instructional tool,''
Am. J. Phys. {\bf 63}, 39--43 (1995).

\bibitem{bachorGEQO} 
H. A. Bachor and T. C. Ralph, {\it A Guide to Experiments in Quantum Optics} (Wiley-VCH, Weinheim, 2004).

\bibitem{prlentangtwinopo} 
A. S. Villar, L. S. Cruz, K. N. Cassemiro, M. Martinelli, and P. Nussenzveig, 
``Generation of bright two-color continuous variable entanglement,'' 
Phys. Rev. Lett. {\bf 95}, 243603-1--4 (2005). 

\bibitem{pdh_ajp01}
E. D. Black,
``An introduction to Pound-Drever-Hall laser frequency stabilization,''
Am. J. Phys. {\bf 69}, 79--87 (2001).

\bibitem{pdhoriginal_applphysb83}
R. W. P. Drever, J. L. Hall, F. V. Kowalski, J. Hough, and G. M. Ford,
``Laser phase and frequency stabilization using an optical resonator,''
Appl. Phys. B {\bf 31}, 97--105 (1983).

\bibitem{levenson_pra85}
M. D. Levenson, R. M. Shelby, A. Aspect, M. Reid and D. F. Walls,
``Generation and detection of squeezed states of light by nondegenerate four-wave mixing in an optical fiber,''
Phys. Rev. A {\bf 32}, 1550--1562 (1985).

\bibitem{galatola_optcomm91}
P. Galatola, L. A. Lugiato, M. G. Porreca, P. Tombesi, and G. Leuchs,
``System control by variation of the squeezing phase,''
Opt. Commun. {\bf 85}, 95--103 (1991).

\bibitem{squeezeprimer_ajp88}
R. W. Henry and S. C. Glotzer,
``A squeezed-state primer,''
Am. J. Phys. {\bf 56}, 318--328 (1988).

\bibitem{ekertknightsqueeze_ajp89}
A. K. Ekert and P. L. Knight,
``Correlations and squeezing of two-mode oscillations,''
Am. J. Phys. {\bf 57}, 692--697 (1989).

\bibitem{qucakes_ajp00}
P. G. Kwiat and L. Hardy,
``The mystery of the quantum cakes,''
Am. J. Phys. {\bf 68}, 33--36 (2001).

\bibitem{fabresemiclassical_leshouches92}
C. Fabre and S. Reynaud, 
``Quantum noise in optical systems: A semiclassical approach,''
in Les Houches Session LIII, edited by J. Dalibard, J. M. Raimond, and J. Zinn-Justin 
(Elsevier, New York, 1992), pp. 675--711.

\bibitem{foot1}This concept fails for very weak 
fields, because a phase operator is a difficult object to define rigorously.\cite{mandelwolfOCQO}

\bibitem{schleich_nature87}
W. Schleich and J. A. Wheeler,
``Oscillations in photon distribution of squeezed states and interference in phase space,''
Nature {\bf 326}, 574--577 (1987).

\bibitem{wallsmilburnQO} 
D. F. Walls and G. J. Milburn, {\it Quantum Optics} (Springer-Verlag, Berlin, 2007).

\bibitem{mandelwolfOCQO} 
L. Mandel and E. Wolf, {\it Optical Coherence and Quantum Optics} (Cambridge University Press, Cambridge, 1995).

\bibitem{gardinerQN} C. W. Gardiner, {\it Quantum Noise} (Springer-Verlag, Berlin, 1991).

\bibitem{levenson_applphysb83}
G. C. Bjorklund, M. D. Levenson, W. Lenth, and C. Ortiz,
``Frequency modulation (FM) spectroscopy,''
Appl. Phys. B {\bf 32}, 145--152 (1983).

\bibitem{pengcavtomo_josab00}
J. Zhang, T. Zhang, K. Zhang, C. Xie, and K. Peng, 
``Quantum self-homodyne tomography with an empty cavity,''
J. Opt. Soc. Am. B {\bf 17}, 1920--1925 (2000).

\bibitem{giacomellicavtomo_pra02}
A. Zavatta, F. Marin, and G. Giacomelli, 
``Quantum-state reconstruction of a squeezed laser field by self-homodyne tomography,''
Phys. Rev. A {\bf 66}, 043805-1--8 (2002).

\bibitem{corresponds}Note that $L$ corresponds to twice the distance between the mirrors in the specific case of a linear cavity (Fig.~\ref{cavotica}). This notation is employed to easily allow for any cavity geometry.

\bibitem{noise}The definition of the noise power involves the variance of the fluctuations (the ellipse). It has been tacitly assumed that the field state possesses a Gaussian Wigner function (the fluctuations satisfy Gaussian statistics), because it is the common situation found in the laboratory. Higher order moments necessary to reconstruct the state in the more general case are in principle accessible from the direct measurement of the fluctuations, as shown by Eq.~(\ref{pr}), given that the linearization procedure of Eq.~(\ref{defdeit}) applies. The statistics of any quadrature direction in phase space can be completely transferred (disregarding losses) to the reflected amplitude statistics.

\end{thebibliography}
\end{document}